\documentclass[floatfix,reprint,amsmath,amssymb,aps,pra]{revtex4-1}
\usepackage{times,mathptmx}
\usepackage{graphicx}
\usepackage[english]{babel}
\usepackage{epstopdf}
\usepackage{xcolor}

\begin{document}

\title{Nonlocal homogenization of $\mathcal{PT}$-symmetric multilayered structures}

\author{Denis~V.~Novitsky$^{1,2}$}
\email{dvnovitsky@gmail.com}
\author{Alexander~S.~Shalin$^{2,3}$}
\author{Andrey~Novitsky$^{4}$}
\email{andreyvnovitsky@gmail.com}

\affiliation{$^1$B.~I.~Stepanov Institute of Physics, National
Academy of Sciences of Belarus, Nezavisimosti Avenue 68, 220072
Minsk, Belarus \\ $^2$ITMO University, Kronverksky Prospekt 49,
197101 St. Petersburg, Russia \\ $^3$Ulyanovsk State University, Lev Tolstoy Street 42, 432017 Ulyanovsk, Russia \\ $^4$Department of Theoretical Physics and Astrophysics, Belarusian State University, Nezavisimosti Avenue 4, 220030 Minsk, Belarus}

\date{\today}

\begin{abstract}
Unique and highly tunable optical properties of $\mathcal{PT}$-symmetric systems and metamaterials enable a plenty of entirely new linear and nonlinear optical phenomena with numerous applications, e.g., for designing subdiffraction lenses, nonreciprocal devices, etc. Therefore, the artificial media with the $\mathcal{PT}$ symmetry attract ever-increasing attention and are now a subject for intensive investigations. One of the commonly used methods providing information about the optical response of artificial nanostructural media is a so called effective medium theory. Here, we examine the possibility of utilizing the effective medium theory for a comprehensive analysis of $\mathcal{PT}$-symmetric multilayered systems composed of alternating loss and gain slabs. We show that applicability of local effective material parameters (or Maxwell Garnett approximation) is very limited and cannot be exploited for a prediction of exceptional points marking a $\mathcal{PT}$ symmetry breaking. On the other hand, nonlocal bianisotropic effective medium parameters can be reliably used, if the thickness of a unit cell is much smaller than the radiation wavelength. In the case of obliquely incident plane waves, we reveal the limitation on the loss-gain coefficient, which should not be too large compared with the real part of the permittivity. We believe that our findings can improve the fundamental understanding of physics behind $\mathcal{PT}$-symmetric systems and advance the development of auxiliary tools for analyzing their peculiar optical response. 
\end{abstract}

\maketitle

\section{Introduction}

$\mathcal{PT}$ symmetry entered physics as a realization of the
non-Hermitian quantum mechanics keeping eigenvalues real \cite{Bender1998,Bender2007}. It soon turned out that the properties of $\mathcal{PT}$-symmetric systems can be relatively easily proved in practice in the optical \cite{El-Ganainy2007,Ruter2010} and microwave \cite{Liu2018} domains (see also the recent review articles \cite{Feng2017,El-Ganainy2018}).
Generally, a $\mathcal{PT}$-symmetric optical system is a periodic structure with
balanced loss and gain (the same value of loss and gain coefficients): the
field of the eigenmodes is equally distributed between the loss and
gain components. When the loss-gain coefficient increases, the balance
may spontaneously disappear at the point of symmetry breaking
(exceptional point), where the eigenstates become degenerate. In the
non-$\mathcal{PT}$-symmetric regime, the field is asymmetrically distributed
between the loss and gain parts of the system producing amplifying
and decaying eigenmodes. The system may return to the $\mathcal{PT}$-symmetric
state at the following exceptional point (see, e.g., Ref. \cite{Novitsky2018}).

Optical $\mathcal{PT}$ symmetry is the basic concept for various prospective
applications including lasing \cite{Feng2014,Hodaei2014} and coherent perfect absorption \cite{Longhi2010,Wong2016}, enhanced sensing
\cite{Chen2017,Hodaei2017,Chen2018}, effects of asymmetric light propagation such as unidirectional invisibility
\cite{Lin2011}, nonreciprocity \cite{Makris2008}, localization \cite{Ji2018}, and others. Optical $\mathcal{PT}$-symmetric systems have usually either a waveguide or multilayer configuration, although there are more exotic variants, such as metasurfaces \cite{Fleury2014, Lawrence2014} or graphene-based structures \cite{Sarisaman2018}. $\mathcal{PT}$ waveguides can be described in the
coupled-mode approximation and can be both active \cite{Guo2009}
and passive \cite{Feng2013}. The multilayer structure as an important system to
understand basics of the PT symmetry without using any approximations will be analyzed in this paper.

A multilayer can be considered as a photonic crystal or a simplest
metamaterial (artificial periodic subwavelength structure) \cite{Shalin2015, Chebykin2015,Slobozhanyuk2015} depending on the relation between the size of the unit cell and the radiation wavelength. A key problem of the metamaterial theory is a homogenization that allows us to treat the metamaterial as a quasicontinuous medium with a set of effective parameters, such as dielectric permittivity and magnetic permeability. Metamaterials of the multilayer geometry can be homogenized using the standard techniques, such as first-principle homogenization \cite{Alu2012,Novitsky2012}, nonlocal effective-medium theory
\cite{Liu2013,Ciattoni2015,Popov2016}, Whitney interpolation
\cite{Tsukerman2011}, etc. However, to the best of our knowledge, there has been no thorough investigation of the homogenization of $\mathcal{PT}$-symmetric systems. In the most relevant article \cite{Shramkova2016}, the $\mathcal{PT}$-symmetric system is composed of the hyperbolic metamaterial and gain medium, the
former being described using the Maxwell Garnett approach. Although the accurate transfer-matrix solution is available for multilayered systems, it is instructive to have a homogenized solution, too. It may not only simplify the description, but also unveil novel regularities. For example, in Ref. \cite{Popov2016} the nonlocal homogenization is used to
derive the criterion of the effective medium theory breakdown. In Ref. \cite{Popov2018}, the nonlocal homogenization theory is exploited to prove that the Maxwell Garnett approach is more applicable for the unit cells with inversion symmetry compared to the unit cells without it.

In this paper, we employ the operator effective medium approximation
(OEMA) \cite{Liu2013,Popov2016} to investigate its area of validity
in description of $\mathcal{PT}$-symmetric multilayered systems. OEMA juxtaposes
a homogeneous nonlocal bianisotropic effective medium to the
multilayer, thus allowing us to accurately find the
transmission and reflection spectra \cite{Popov2016} and surface-wave
propagation \cite{Popov2018}. Neglecting the nonlocal effects, the
OEMA provides the Maxwell Garnett approximation and the well-known
mixing formulas for the components of the effective permittivity tensor
\cite{Markel2016}. In Section II, we write out the effective medium tensors derived in Ref. \cite{Popov2016} for the $\mathcal{PT}$-symmetric multilayer in the zeroth, first, and second orders of the OEMA. In Section III, we discuss
the transmission and reflection characteristics and description of
$\mathcal{PT}$ symmetry breaking using the local and nonlocal material
parameters and find the limits of their applicability. In Section IV, we generalize the obtained regularities to the two-dimensional $\mathcal{PT}$-symmetric systems. Section V sums up the main results of the article.

\section{Operator effective medium approximation for a $\mathcal{PT}$-symmetric multilayer system}

\begin{figure}[t!]
\centering \includegraphics[scale=0.7, clip=]{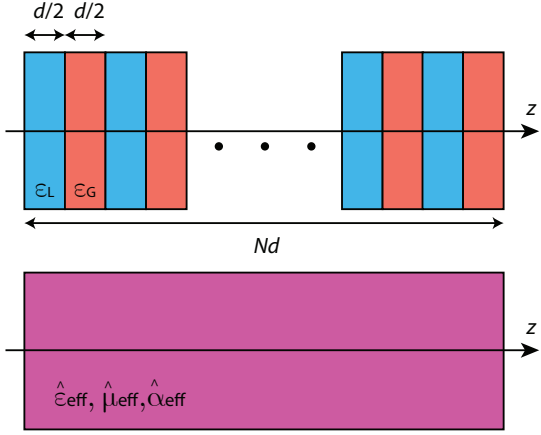}
\caption{\label{fig1} Geometry of the $\mathcal{PT}$-symmetric system as an
$N$-unit-cell multilayer composed of alternating loss and gain slabs
of equal thickness $d/2$ and permittivities $\varepsilon_L =
\varepsilon' + i \varepsilon''$ and $\varepsilon_G = \varepsilon' -
i \varepsilon''$, respectively. A homogeneous bianisotropic slab with
material tensors $\hat\varepsilon_{eff}$, $\hat\mu_{eff}$, and
$\hat\alpha_{eff}$ calculated using the OEMA is shown in the bottom part
of the sketch.}
\end{figure}

We consider a $\mathcal{PT}$-symmetric multilayered structure illustrated in Fig.
\ref{fig1}. It consists of alternating loss and gain layers of the
same thickness $d/2$, the total number of layers being $2N$. To
ensure the optical $\mathcal{PT}$ symmetry, the permittivity in the multilayer
should be distributed as $\varepsilon (z) = \varepsilon^\ast (z)$, which is
realized using the loss $\varepsilon_L=\varepsilon' + i
\varepsilon''$ and gain $\varepsilon_G=\varepsilon' - i
\varepsilon''$ permittivities (loss-gain coefficient
$\varepsilon''>0$). Such a simple structure can be fully described
using the transfer-matrix method (TMM), including its transmission and
reflection (scattering) properties. We will exploit the TMM for
homogenized slabs as well.

To homogenize the multilayered system, we use a recently developed
operator effective medium approximation \cite{Liu2013,Popov2016}.
The idea behind the OEMA is based on writing the fundamental
solution for the layered periodic structure as a series over the
size parameter $k_0 d$, where $k_0$ is the vacuum wavenumber and $d$
is the thickness of the unit cell. Then one is able to introduce the
$m$-th order of approximation as a solution containing the terms up
to $(k_0 d)^m$. The larger the order, the better the approximation.
We do not derive equations for the effective material parameters
here, but just borrow them from previous research
\cite{Popov2016}. Let us start with the zeroth-order ($m=0$)
approximation known as the Maxwell Garnett approximation.

\subsection{Maxwell Garnett approximation}

The Maxwell Garnett approach is believed to be valid, if $k_0 d \ll 1$.
In this case, the loss/gain multilayer can be represented as a
homogeneous uniaxial medium characterized by the permittivity tensor
\begin{eqnarray}
\hat\varepsilon^{(0)} &=& \left( \begin{array}{ccc}
\varepsilon^{(0)}_{||} & 0 & 0 \\ 0 & \varepsilon^{(0)}_{||} & 0 \\
0 & 0 & \varepsilon^{(0)}_{\perp} \end{array} \right), \label{MaxGar} \\
\varepsilon^{(0)}_{||}&=&\frac{\varepsilon_L
d/2 + \varepsilon_G d/2}{d} = \varepsilon', \nonumber \\
\varepsilon^{(0)}_{\perp}&=&\frac{d}{\varepsilon_L^{-1} d/2 + \varepsilon_G^{-1} d/2} = \frac{\varepsilon^{\prime
2}+\varepsilon^{\prime\prime 2}}{\varepsilon'}. \nonumber 
\end{eqnarray}
Since the size parameter enters the above expression as
$(k_0 d)^0$, the Maxwell Garnett approach is the zeroth-order
approximation. It should be stressed that this oversimplified
technique may not be satisfactory, even if $k_0 d \ll 1$ (see the
works on the breakdown of the effective medium theory, e.g., Refs. 
\cite{Sheinfux2014,Zhukovsky2015}).

\subsection{First-order OEMA}

Material parameters in the first-order approximation imply nonlocality
and bianisotropy. Bianisotropic (magnetoelectric) terms emerge in
the first order, $(k_0 d)^1$, together with the permittivity tensor
(\ref{MaxGar}). The magnetoelectric coupling tensor was derived in Ref.
\cite{Popov2016} and for the loss/gain multilayered system equals
\begin{eqnarray}
\hat\alpha &=& \left( \begin{array}{ccc} 0 & \alpha_1 & 0 \\
\alpha_2 & 0 & 0 \\ 0 & 0 & 0 \end{array} \right), \label{MEcoupl} \\
\alpha_1 &=& \frac{\varepsilon'' k_0 d}{4}, \qquad \alpha_2 =
\frac{\varepsilon'' k_0 d}{4} \left[ \frac{2 \varepsilon'
k_t^2}{k_0^2 (\varepsilon^{\prime 2} + \varepsilon^{\prime \prime
2})} -1 \right]. \nonumber
\end{eqnarray}
Here the nonlocality appears as evidenced by the dependence of the material parameters on
the tangential wavenumber $k_t$, i.e., the projection of the wavevector on
the interface between slabs. The constitutive equations for such a
bianisotropic medium read ${\bf D} = \hat\varepsilon^{(0)} {\bf E} +
\hat\alpha {\bf H}$ and ${\bf B} = {\bf H} + \hat\alpha^T {\bf E}$,
where ${\bf E}$ and ${\bf H}$ (${\bf D}$ and ${\bf B}$) are the
electric and magnetic field strengths (inductions) and superscript
$T$ stands for the transposition. It is worth noticing that the
magnetoelectric coupling (\ref{MEcoupl}) depends on the wave
propagation direction, that is, the order of layers ($\varepsilon_L
\leftrightarrow \varepsilon_G$ is equivalent to $\varepsilon''
\leftrightarrow -\varepsilon''$).

\subsection{Second-order OEMA}

The second-order corrections influence the effective permittivity
and permeability tensors keeping the effective magnetoelectric
coupling tensors as defined according to Eq.
(\ref{MEcoupl}). According to Ref. \cite{Popov2016}, the
permittivity and permeability tensors for the loss-gain multilayer
take the form
\begin{eqnarray}
\hat\varepsilon^{(2)} &=& \left( \begin{array}{ccc}
\varepsilon^{(2)}_{||} & 0 & 0 \\ 0 & \varepsilon^{(2)}_{||} & 0 \\
0 & 0 & \varepsilon^{(2)}_{\perp} \end{array} \right), \quad
\hat\mu^{(2)}= \left( \begin{array}{ccc}
\mu^{(2)}_{||} & 0 & 0 \\ 0 & \mu^{(2)}_{||} & 0 \\
0 & 0 & \mu^{(2)}_{\perp} \end{array} \right), \nonumber \\
\varepsilon^{(2)}_{||} &=& \varepsilon^{(0)}_{||}(1+ w), \qquad
\varepsilon^{(2)}_{\perp}=\varepsilon^{(0)}_{\perp}(1- w), \nonumber
\\
\mu^{(2)}_{||} &=& 1, \qquad \mu^{(2)}_{\perp} = 1- \frac{ (k_0
d)^2}{6} \frac{\varepsilon' \varepsilon^{\prime \prime
2}}{\varepsilon^{\prime 2} + \varepsilon^{\prime \prime 2}},
\nonumber \\
w &=& \frac{(k_0 d)^2 \varepsilon^{\prime \prime 2}}{12
\varepsilon^{\prime 2}} \left[ \frac{2 \varepsilon' k_t^2}{k_0^2
(\varepsilon^{\prime 2} + \varepsilon^{\prime \prime 2})} -1
\right]. \label{OEMA2or}
\end{eqnarray}

The permittivity correction is electric quadrupolar and nonlocal
($k_t$-dependent). The permeability is caused by the artificial
magnetic moment due to the displacement currents. The effective
medium in the second-order OEMA is characterized by the constitutive
equations ${\bf D} = \hat\varepsilon^{(2)} {\bf E} + \hat\alpha {\bf
H}$ and ${\bf B} = \mu^{(2)} {\bf H} + \hat\alpha^T {\bf E}$.

Homogenized PT-symmetric systems are characterized by the peculiar material parameters. They are real-valued quantities, while the permittivity and permeability tensors do not depend on the sign of $\varepsilon''$. Next we will study whether the homogenization is able to predict the positions of exceptional points in $\mathcal{PT}$-symmetric structures.

\section{Wave propagation in homogenized $\mathcal{PT}$-symmetric slab}

\subsection{Normal incidence}

\begin{figure}[t!]
{\includegraphics[scale=1., clip=]{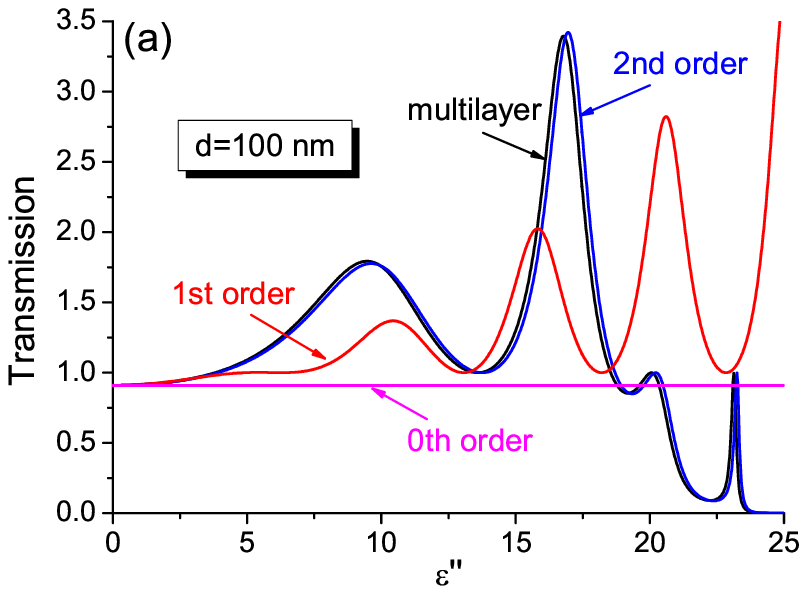}
\includegraphics[scale=1., clip=]{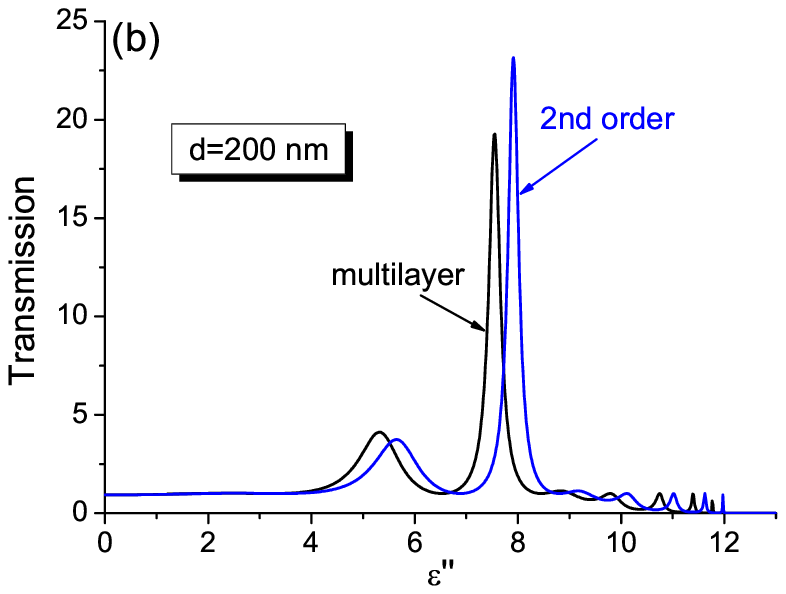}}
\caption{\label{fig2} Transmission of the $\mathcal{PT}$-symmetric multilayer as a
function of $\varepsilon''$ calculated within the full TMM (indicated with ``multilayer'') and
OEMA of different orders for the period thickness (a) $d=100$ nm and (b) $d=200$ nm. The case
of a normally incident wave with $\lambda=1.55$
$\mu$m is considered; the structure consists of $N=20$ slabs; $\varepsilon'=2$.
}
\end{figure}

\begin{figure}[t!]
{\includegraphics[scale=1., clip=]{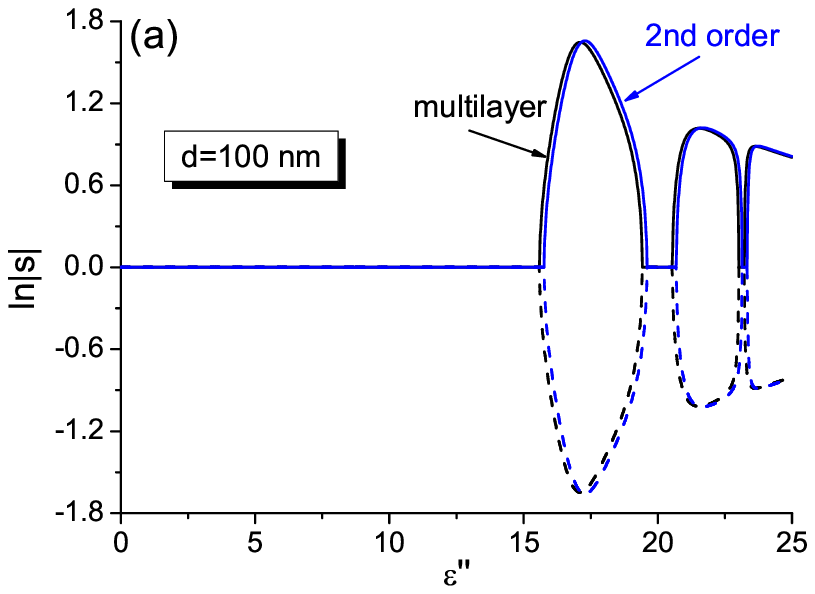}
\includegraphics[scale=1.05, clip=]{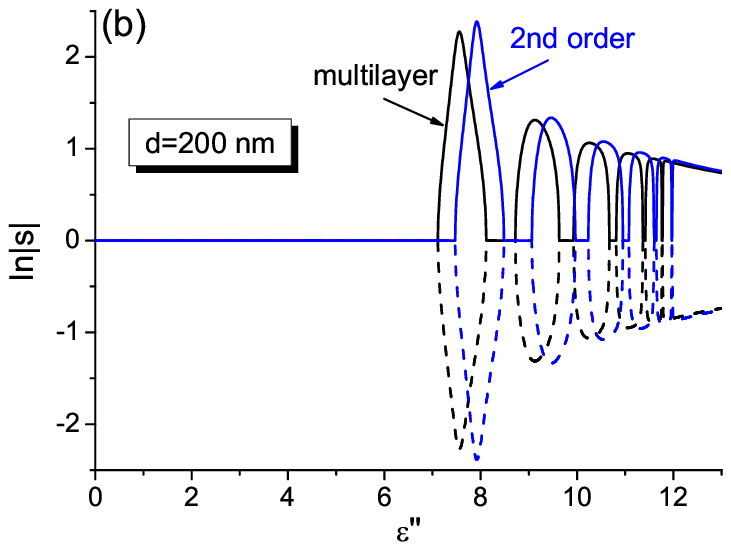}}
\caption{\label{fig3} Dependence of the scattering-matrix
eigenvalues on $\varepsilon''$ calculated within the full TMM
(indicated by ``multilayer'') and
OEMA of different orders for the period thickness (a) $d=100$ nm and (b) $d=200$ nm. Other
parameters are the same as in Fig. \ref{fig2}. The first and the
second eigenvalues are designated with solid and dashed curves,
respectively.}
\end{figure}

\begin{figure}[t!]
\centering \includegraphics[scale=0.95, clip=]{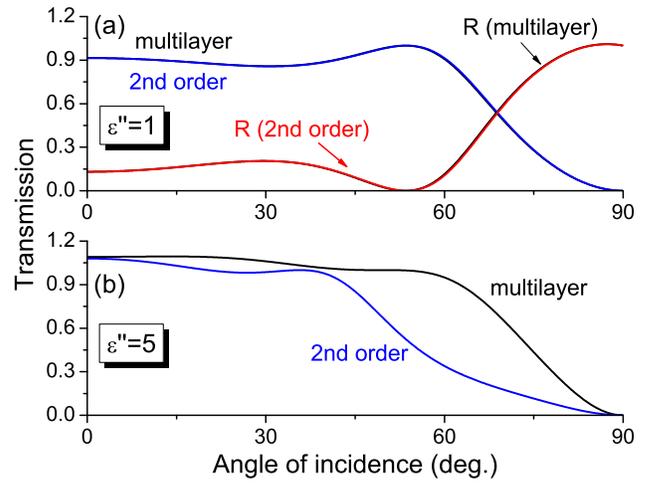}
\caption{\label{fig4} Angular dependence of transmission for the
$\mathcal{PT}$-symmetric multilayer calculated within the full TMM (indicated by ``multilayer'') and the
second-order OEMA. Upper panel shows the results for
$\varepsilon''=1$, lower panel -- for $\varepsilon''=5$. Behavior of
reflection is shown only for $\varepsilon''=1$ and is omitted in the
lower panel not to encumber the figure. The period thickness is $d=100$
nm, and the other parameters are the same as in Fig. \ref{fig2}.}
\end{figure}

\begin{figure}[t!]
\centering \includegraphics[scale=0.95, clip=]{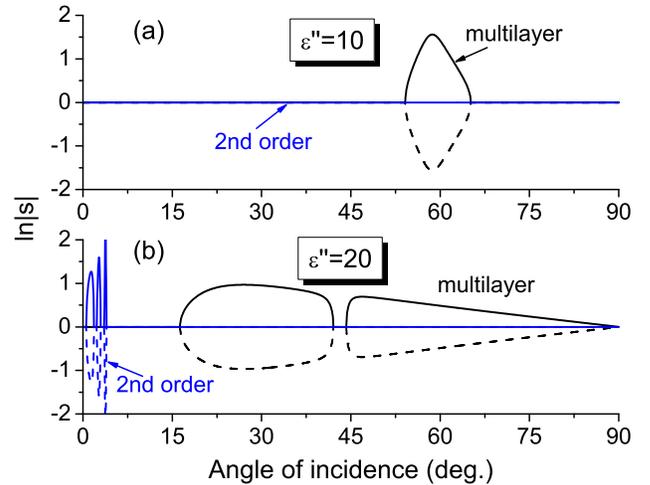}
\caption{\label{fig5} Angular dependence of the scattering-matrix
eigenvalues calculated within the full TMM (indicated by ``multilayer'') and the second-order
OEMA. The upper panel shows the results for $\varepsilon''=10$, the lower one -- for $\varepsilon''=20$. The other parameters are the same as in Fig. \ref{fig4}. The first and the second eigenvalues are designated
with solid and dashed curves, respectively.}
\end{figure}

\begin{figure}[t!]
\centering \includegraphics[scale=1., clip=]{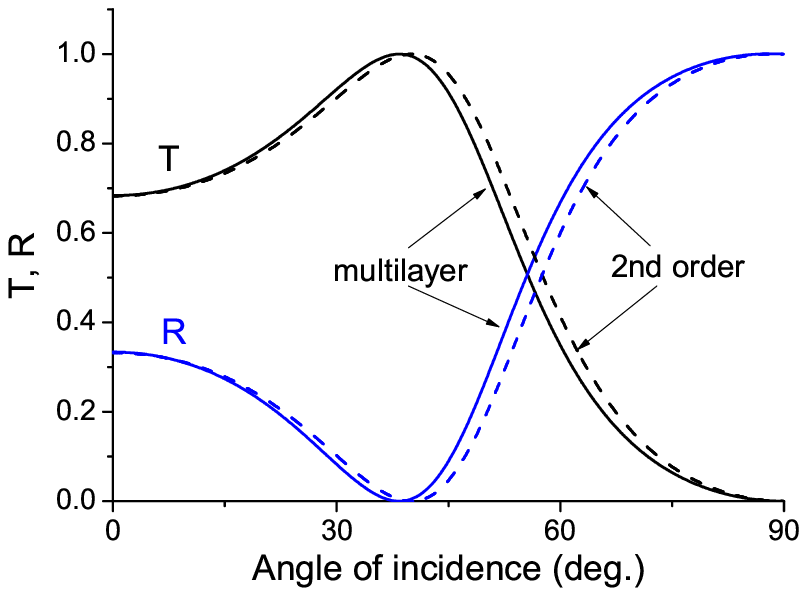}
\caption{\label{fig6} Angular dependence of transmission and
reflection calculated within the full TMM (indicated by ``multilayer'') and the second-order
OEMA for $\varepsilon'=10$ and $\varepsilon''=1$. The other parameters
are the same as in Fig. \ref{fig4}.}
\end{figure}

Here we compare the scattering properties of homogenized versus
multilayered $\mathcal{PT}$-symmetric media. Strictly speaking, the homogenized
slab cannot be treated as the $\mathcal{PT}$-symmetric system, but a wave
propagation in both systems can be quite similar. Homogenized
material is considered as anisotropic (Maxwell Garnett
approximation) or nonlocal bianisotropic (first and second order of
OEMA) media. In Fig. \ref{fig2}, we demonstrate the transmission $T$
of a normally incident plane wave. In the zeroth-order OEMA, the
transmission of the normally incident wave is defined by the
in-plane permittivity $\varepsilon^{(0)}_{||}$, which is why it does
not depend on the imaginary part of permittivity $\varepsilon''$
[see the horizontal straight line in Fig. \ref{fig2}(a)]. It should be noticed
that the Maxwell Garnett approach reproduces the correct value of
$T$ only for small $\varepsilon''$, when the second-order
corrections are negligible. The region of validity of the
first-order OEMA is wider, but it is anyway very limited. In the
second order, the transmission spectra follow the curve
corresponding to the transmission through the multilayer. However,
this approximation is valid only for the thin unit cells like $d=100
{\rm nm} \ll \lambda$. For instance, when we double the thickness of
the unit cell, the second order OEMA only qualitatively describes
the transmission reproducing peaks, but not their positions and
heights [see Fig. \ref{fig2}(b)].

Since the thicknesses of the layers are limited by the validity of
the effective medium theory, the large values of the loss-gain coefficient $\varepsilon''$ should be used for reproducing the symmetry breaking in $\mathcal{PT}$-symmetric
systems, when loss and gain are no longer balanced. We exploit the
standard means for studying the $\mathcal{PT}$ symmetry breakdown, i.e., the
calculation of eigenvalues and eigenvectors of the system's scattering
matrix. The scattering matrix for the multilayered system and the
homogenized medium under consideration can be derived from the
corresponding transfer matrix \cite{Ge2012}. The eigenvalues of the scattering matrix are unitary
($|s_1|=|s_2|=1$) in the $\mathcal{PT}$-symmetric state and inverse
($|s_1|=1/|s_2|>1$) in the $\mathcal{PT}$-broken state. The eigenvectors of the
scattering matrix coalesce at the exceptional point (point of the non-Hermitian
singularity), where a phase transition from the $\mathcal{PT}$-symmetric to the
non-PT-symmetric state occurs.

Let us examine, whether the local or nonlocal homogenization is able
to catch the exceptional points of the $\mathcal{PT}$ symmetry breaking.
Scattering matrix, transmission, and reflection are expected to be
simultaneously well predicted, since the scattering matrix is
defined in terms of the amplitude transmission and reflection
coefficients. In Fig. \ref{fig3}(a), we show dependence of the
eigenvalues of the scattering matrix on $\varepsilon''$ for $d=100$
nm. Similar to the transmission curves reported in Fig.
\ref{fig2}(a), the second-order OEMA correctly describes behavior of
the eigenvalues and reproduces the breaking of the $\mathcal{PT}$ symmetry observed in
the TMM calculations of the multilayer. The second-order OEMA is also well
suited for describing the coalescence of the eigenvectors (not
shown here).

Detailed analysis shows that neither the first-order nor zeroth-order
OEMA gives a hint of the phase transition. The range of applicability for
the Maxwell Garnett approximation is limited by the small loss-gain
coefficients $\varepsilon''$, where the $\mathcal{PT}$ symmetry breaking does
not occur. The symmetry breaking could be reached for smaller
$\varepsilon''$, if the unit cells were thicker. However, the larger
$d$ would also ruin the Maxwell Garnett approximation. It is worth
mentioning that the homogeneous slab characterized by the
permittivity tensor Eq. (\ref{MaxGar}) is inherently inappropriate for description of the exceptional points, because its transmission
and reflection properties do not depend on the direction of wave
incidence. For instance, in the case of the normal incidence, the properties of the
homogeneous slab are equivalent to those of an isotropic dielectric
slab of permittivity $\varepsilon'$.

The breakdown of the $\mathcal{PT}$ symmetry cannot be also obtained, if we skip the terms proportional to $(k_0 d)^1$ and leave the second-order terms $(k_0 d)^2$. This
means that only the mix of the first-order and second-order terms
properly reproduces the eigenvalues. As previously, the thicker
layers ruin the correspondence between the exact and homogenized
description of the structure [Fig. \ref{fig3}(b)]. The difference
between the accurate and approximate eigenvalues is mainly the shift
along the $\varepsilon''$-axis. The eigenvalues in the second-order
OEMA $s^{(2)}$ are roughly related to the eigenvalues of the
$\mathcal{PT}$-symmetric multilayer $s$ as $s(\varepsilon'') \approx
s^{(2)}(\varepsilon''+a k_0 d)$, where the shift is proportional to
the size parameter $k_0 d$ and $a$ is a constant. Hence, we can
estimate the higher-order corrections caused by excitation of the
higher-order multipoles as $[d s^{(2)}(\varepsilon'')/d
\varepsilon''] a k_0 d$.

\subsection{Oblique incidence}

Now we face the dependence of the multilayer response on the incidence angle of the plane waves. In Fig. \ref{fig4}, we observe a strong dependence of the
correctness of the homogenization on the value of $\varepsilon''$. For
$\varepsilon''=1$, both transmission and reflection dependencies on
the incidence angle are reliably described by the second-order OEMA,
but this is not already the case at $\varepsilon''=5$. This discrepancy
grows with the increase of $\varepsilon''$: the second-order OEMA
may even predict a number of artifact peaks absent in the curves
calculated for the multilayer.

The validity of the nonlocal effective medium approximation of the
second order is determined by the loss-gain coefficient
$\varepsilon''$, but not the angle of incidence. We relate the
behavior shown in Fig. \ref{fig4}(b) to the appearance of the $\mathcal{PT}$-symmetry-broken states for the larger incidence angles. At the oblique incidence, the light passes a longer path compared to the normal incidence; therefore, the
effective thickness of the structure is enlarged and exceptional
points of the $\mathcal{PT}$ symmetry breaking may appear at the smaller
$\varepsilon''$. In Fig. \ref{fig5} we plot the angular dependencies
of the scattering matrix eigenvalues for $\varepsilon''=10$ and
$\varepsilon''=20$. The results for the multilayered and homogenized structures are
qualitatively different in these cases. The OEMA cannot catch the
$\mathcal{PT}$-symmetry-broken state for $\varepsilon''=10$, because it does not properly
enlarge the effective thickness at the oblique incidence. Indeed,
the effect of the tangential wavenumber $k_t$ is negligible for the
great $\varepsilon''$ [see Eqs. (\ref{MEcoupl}) and
(\ref{OEMA2or})]; i.e., the results are quite close to those for $k_t
= 0$. As well as for the normal incidence, the $\mathcal{PT}$-symmetry-broken
states appear at the larger $\varepsilon''=20$, but they do not
correspond to the accurate calculations.

According to Eqs. (\ref{MEcoupl}) and (\ref{OEMA2or}), the wavenumber
$k_t$ makes significant effect for the greater real part of the
permittivity $\varepsilon'$. Figure \ref{fig6} shows the angular
dependencies of transmission and reflection at $\varepsilon'=10$ and
small loss-gain coefficient $\varepsilon''=1$. Comparing with the
good matching of the accurate and approximate results for
$\varepsilon'=2$ and $\varepsilon''=1$ (see Fig. \ref{fig4}), one
can note that there is some discrepancy between the OEMA and
accurate calculations, but only in the region of large incidence
angles. We conclude that the nonlocal homogenization may be applicable, if
$\varepsilon''$ is not much greater than $\varepsilon'$.

\section{Homogenization of two-dimensional $\mathcal{PT}$-symmetric systems}

\begin{figure}[t!]
\centering \includegraphics[scale=0.7, clip=]{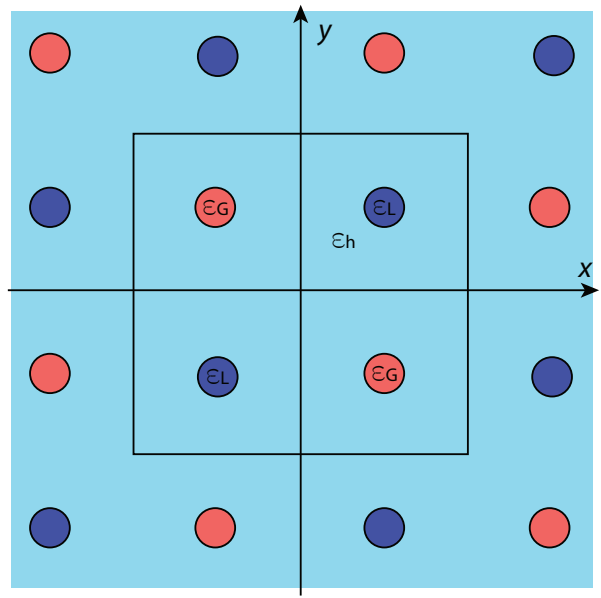}
\caption{\label{fig7} Square unit cell of the two-dimensional $\mathcal{PT}$-symmetric structure comprising four $z$-oriented circular waveguides.}
\end{figure}

Although we have considered only one-dimensional structures above, two-dimensional $\mathcal{PT}$-symmetric systems also can be homogenized. Two-dimensional $\mathcal{PT}$ symmetry can be introduced in a system of periodically arranged $z$-oriented waveguides as in Ref. \cite{Kremer2019}. The system of waveguides can be nonlocally homogenized in different ways. In concordance with a coherent-potential approximation \cite{Wu2006,Slovick2014,Geng2015}, a unit cell of the periodic array should not scatter light, if it is embedded into a proper effective medium. Such a condition can be easily written for a single cylinder, but it is much more complicated for the unit cell in Fig. \ref{fig7} (loss and gain waveguides of the same radius $r$ in the cell). Nevertheless, the effective material parameters in the limit of small cylinder radii $k_0 r \ll 1$ should have a similar form for the single cylinder and four cylinders in the cell. Thus, in the case of the two-dimensional $\mathcal{PT}$-symmetric structure one writes $\hat\varepsilon = {\rm diag}\left( \varepsilon_x, \varepsilon_y, \varepsilon_z \right)$, where \cite{Geng2015}
\begin{equation}
\varepsilon_x= \varepsilon_y= \varepsilon_x^{(0)}, \qquad
\varepsilon_z = \frac{\varepsilon_z^{(0)} + \alpha \rho (\varepsilon_L+\varepsilon_G) k_z^2}{1 + 2\alpha \rho  k_z^2}.\label{epsNL2D}
\end{equation}
Here $\alpha$ is the parameter defined by the cell geometry, and $\rho$ is the filling factor of loss (gain) material. The Maxwell Garnett permittivities are equal to
\begin{eqnarray}
\varepsilon_x^{(0)} &=& \rho (\varepsilon_L + \varepsilon_G) + (1-2\rho) \varepsilon_h, \nonumber \\
\varepsilon_z^{(0)} &=& \varepsilon_h \frac{1+p}{1-p}, \quad p = 2 \rho \frac{\varepsilon_L \varepsilon_G - \varepsilon_h^2}{(\varepsilon_L+\varepsilon_h) (\varepsilon_G+\varepsilon_h)},
\end{eqnarray}
where $\varepsilon_h$ is the host medium permittivity.

The nonlocal permittivity $\varepsilon_z$ in Eq. (\ref{epsNL2D}) depends on the wavenumber $k_z$. It is obtained as a series expansion of the zeroth Mie coefficient for TM-polarized waves \cite{Geng2015}. This means that the nonlocality is related to the toroidal moment \cite{Liu2015}, but not to the quadrupole moment defined by the second Mie coefficient \cite{Novitsky2017}. On the other hand, as we have discussed above, the electric quadrupole moment is needed for predicting exceptional points in $\mathcal{PT}$-symmetric multilayers. Thus, the homogenization carried out using the coherent-potential approximation cannot reproduce exceptional points of $\mathcal{PT}$ symmetry breaking as well as the Maxwell Garnett approximation.

Another approach is the nonlocal homogenization theory, which can be applied to an arbitrary metamaterial. The effective dielectric permittivity tensor of the homogenized medium takes the form \cite{Ciattoni2015}
\begin{equation}
\epsilon_{ij}^{(2)} = \varepsilon_{ij} + i \alpha_{ijr}
k_r - \beta_{ijrs} k_r k_s,
\label{eps2D}
\end{equation}
where the summation over repeated indices is employed, indices $i$, $j$, $r$, and $s$ run from 1 to 3, and $k_r$ is the $r$-component of the wavevector. Tensors in Eq. (\ref{eps2D}) depend on distribution of the permittivity in the metamaterial $\varepsilon(\textbf{r}) = \varepsilon'(\textbf{r})+i \varepsilon''(\textbf{r})$ and expansion parameter $k_0 d$, being symmetric $\epsilon_{ij} =  \epsilon_{ji} =  \xi_{ij}^{(1)} + (k_0 d)^2  \xi_{ij}^{(2)}$, $\beta_{ijrs} = \beta_{jirs} = (k_0 d)^2 \gamma_{ijrs}$ or antisymmetric $\alpha_{ijr} = -\alpha_{jir} = (k_0 d) \zeta_{ijr}$. The condition for $\mathcal{PT}$ symmetry $\varepsilon(\textbf{r}) = \varepsilon^*(-\textbf{r})$ imposes the following restrictions on the tensors: $\epsilon_{ij} =  \epsilon_{ij}^*$, $\alpha_{ijr} = -\alpha_{ijr}^*$, and $\beta_{ijrs} = \beta_{ijrs}^*$. The Maxwell Garnett approximation corresponds to $\varepsilon_{ij}^{(0)} = \xi_{ij}^{(1)}$. The structure of Eq. (\ref{eps2D}) is similar to that of the periodic multilayer. In both approaches, the zeroth order corresponds to the Maxwell Garnett approximation, the first order introduces gyrotropy (chirality), and the second order takes into account quadrupole moment. Moreover, in Ref. \cite{Ciattoni2015} the general nonlocal homogenization theory was applied for the stack of layers. Therefore, Eq. (\ref{eps2D}) is an ansatz for both one- and two-dimensional $\mathcal{PT}$-symmetric systems and, hence, prediction of the exceptional points should be valid in the general case. Homogenization of two-dimensional $\mathcal{PT}$-symmetric systems in details deserves separate investigation.

\section{Conclusion}

To conclude, we have studied the local and nonlocal homogenization of the
$\mathcal{PT}$-symmetric multilayered system. It has been carried out using the
previously developed operator effective medium approximation
providing the successive approximations for effective parameters including the usual mixing
(Maxwell Garnett) formulas and the nonlocal bianisotropic material
tensors. We have found that the Maxwell Garnett approach is entirely
inappropriate for description of the $\mathcal{PT}$ symmetry. The nonlocal model
takes into account the distribution of the loss and gain materials
and can be applied in the limit of electrodynamically small
thicknesses of the unit cells. Such behavior is equally related to
the transmission and scattering matrix' eigenvalues spectra. For the
obliquely incident plane waves, the validity of the nonlocal
homogenization is limited by the loss-gain coefficients
$\varepsilon''$ comparable to or smaller than the real part of the
permittivity $\varepsilon'$. Thus, though the nonlocal homogenization is applicable, it is strongly restricted by the thickness of the unit cell and the value of the loss-gain coefficient.

\acknowledgements{The work was supported by the Belarusian
Republican Foundation for Fundamental Research (Project No.
F18R-021), the Russian Foundation for Basic Research (Projects No.
18-02-00414, No. 18-52-00005 and No. 18-32-00160), the Ministry of
Education and Science of the Russian Federation (GOSZADANIE, Grant
No. 3.4982.2017/6.7), and the Government of the Russian Federation
(Grant No. 08-08). Calculations of transmission and reflection from periodic media were supported by the Russian Science Foundation (Project No. 17-72-10098).}

\end{document}